\documentstyle[12pt]{article}
\begin{document}
\begin{titlepage}
\pagestyle{empty}
\baselineskip=21pt
\begin{center}
{\large{\bf Predictions from Quantum Cosmology}} \footnote {Lectures
at International School of Astrophysics ``D.Chalonge'', Erice, 1995.}
\end{center}
\vskip .1in
\begin{center}
Alexander Vilenkin

{\it Institute of Cosmology, Department of Physics and Astronomy}

{\it Tufts University, Medford, MA 02155, USA}

\vskip .1in

\end{center}
\vskip .5in
\centerline{ {\bf Abstract} }
\baselineskip=18pt

After reviewing the general ideas of quantum cosmology (Wheeler-DeWitt
equation, boundary conditions, interpretation of $\psi$), I discuss
how these ideas can be tested observationally.  Observational
predictions differ for different choices of boundary conditions.  With
tunneling boundary conditions, $\psi$ favors initial states that lead
to inflation, while with Hartle-Hawking boundary conditions it does
not.  This difficulty of the Hartle-Hawking wave function becomes
particularly severe if the role of `inflatons' is played by the moduli
fields of superstring theories.  In models where the constants of
Nature can take more than one set of values, $\psi$ can also determine
the probability distribution for the constants.  This can be done with
the aid of the `principle of mediocrity' which asserts that we are a
`typical' civilization in the ensemble of universes described by
$\psi$.  The resulting distribution favors inflation with a very flat
potential, thermalization and baryogenesis at electroweak scale,
a non-negligible cosmological constant, and density fluctuations seeded either
by topological defects, or by quantum fluctuations in models like hybrid
inflation (as long as these features are
consistent with the allowed values of the constants).

\noindent

\end{titlepage}
\newpage
\baselineskip=18pt

\noindent{\bf 1.\quad Introduction}
\medskip
\nobreak

If the cosmological evolution is followed back in time, we come to the
initial singularity where the classical equations of general
relativity break down.  This led many people to believe that in order
to understand what actually happened at the origin of the universe, we
should treat the universe quantum-mechanically and describe it by a
wave function rather than by a classical spacetime.  This quantum
approach to cosmology was initiated by DeWitt \cite{Dewitt} and Misner
\cite{Misner}, and after a somewhat slow start has become very popular
in the last decade or so.  The picture that has emerged from this line
of development \cite{AV82,HH83,Zelstar,Linde84,Rub84,AV84,Tryon} is
that a small closed universe can spontaneously nucleate out of
nothing, where by `nothing' I mean a state with no classical space and
time.  The cosmological wave function can be used to calculate the
probability distribution for the initial configurations of the
nucleating universes.  Once the universe nucleated, it is expected to
go through a period of inflation, which is a rapid (quasi-exponential)
expansion driven by the energy of a false vacuum.  The vacuum energy
is eventually thermalized, inflation ends, and from then on the
universe follows the standard hot cosmological scenario.  Inflation is
a necessary ingredient in this kind of scheme, since it gives the only
way to get from the tiny nucleated universe to the large universe we
live in today.

Another possible use for quantum cosmology is to determine the
probability distribution for the values of the constants of Nature.
The constants can vary from one universe to another due to a different
choice of the vacuum state, a different compactification scheme in
higher-dimensional theories, or to Planck-scale wormhole effects
\cite{Coleman}.  The cosmological wave function will then be a
superposition of terms corresponding to all possible values of the
constants.

In these lectures, I would like to review where we stand in this program.
The general ideas of quantum cosmology and predictions for the initial
state are discussed in Section 2-4, followed by a discussion of
predictions for the constants of Nature in Sections 5,6.
Due to the time constraints, some important topics will be left out.
These include topology-changing processes, third quantization,
consistent histories approach, and decoherence.
\bigskip
\newpage
\baselineskip=18pt

\noindent{\bf 2.\quad A Simple Model}

\smallskip

\noindent{2.1\quad `THE GREATEST MISTAKE OF MY LIFE'}
\medskip
\nobreak

First I would like to illustrate how the nucleation of a universe can be
described in a very simple model.  The model is defined by the action
\begin{equation}
S = \int d^4x\sqrt{-g}\left({R\over{16\pi G}}-\rho_v\right),
\label{action}
\end{equation}
where $\rho_v$ is a constant vacuum energy and the universe is assumed
to be homogeneous, isotropic, and closed,
\begin{equation}
ds^2 =\sigma^2 [-dt^2 + a^2 (t)d\Omega_3^2 ].
\label{metric}
\end{equation}
Here, $d\Omega_3^2$ is the metric on a unit three-sphere, and
$\sigma^2 = 2G/3\pi$ is a normalizing factor chosen for later
convenience.  The scale factor $a(t)$ satisfies the evolution equation
\begin{equation}
{\dot a}^2 +1 -H^2a^2 =0,
\label{evoleq}
\end{equation}
where
\begin{equation}
H=4G\rho_v^{1/2} /3.
\end{equation}
The solution of Eq.(\ref{evoleq}) is the de Sitter space,
\begin{equation}
a(t)=H^{-1}\cosh (Ht).
\label{desitter}
\end{equation}
The universe contracts at $t<0$, reaches the minimum radius
$a=H^{-1}$ at $t=0$, and re-expands at $t>0$.

This is similar to the behavior of a particle bouncing off a potential
barrier, with $a$ playing the role of particle coordinate.  Now, we
know that in quantum mechanics particles can not only bounce off, but
can also tunnel through potential barriers.  This suggests the
possibility that the negative-time part of the evolution in
(\ref{desitter}) may be absent, and that the universe may instead
tunnel from $a=0$ directly to $a=H^{-1}$.

When I suggested this idea in 1982, I made an attempt to estimate the
tunneling probability in the semiclassical approximation.  To describe
the tunneling process, I used the bounce solution of the Euclidean
field equations, which can be obtained by substituting $t=-i\tau$ in
Eqs.(\ref{metric}),(\ref{desitter}),
 \begin{equation}
 ds^2=\sigma^2 [d\tau^2 +{\tilde a}^2 (\tau)d\Omega_3^2 ],
 \end{equation}
 \begin{equation}
 {\tilde a}(\tau)=H^{-1}\cos (H\tau).
 \label{bounce}
 \end{equation}
 This metric describes a four-sphere $S^4$ of radius $H^{-1}$.  The
nucleation of the universe is schematically represented in Fig.~1,
where the bounce solution (\ref{bounce}) connects to the Lorentzian
solution (\ref{desitter}) at the turning point $\tau =t=0$.

\begin{figure}
\vspace{8cm}\caption{\it A schematic representation of the birth of
inflationary universe.}
\end{figure}

For `normal' quantum tunneling, the tunneling probability ${\cal P}$ is
proportional to $\exp (-S_E)$, where $S_E$ is the Euclidean action for
the corresponding bounce.  In our case,
\begin{equation}
S_E = \int d^4 x \sqrt{-g}\left(-{R\over{16\pi G}}+\rho_v\right) =
-2\rho_v \Omega_4\sigma^4 H^{-4} = -3/8G^2\rho_v ,
\label{eucaction}
\end{equation}
where
\begin{equation}
R=12H^2 =32\pi G\rho_v
\end{equation}
is the scalar curvature, and $\Omega_4 = 4\pi^2 /3$ is the volume of a
unit four-sphere.  Hence, I concluded in Ref. \cite{AV82} that
\begin{equation}
{\cal P} \propto \exp \left({3\over{8G^2\rho_v}}\right).
\label{wrongeq}
\end{equation}
Following fashion, I might declare this `the greatest mistake of my
life' \cite{Mistake}.
\bigskip

\noindent{2.2\quad THE TUNNELING WAVE FUNCTION}
\medskip
\nobreak

What I now think is the correct answer is given by ${\cal P} \propto
\exp (-|S_E|)$.  In the case of `normal' quantum tunneling, the
Euclidean action is positive-definite, and $|S_E|=S_E$, but for
quantum gravity this is no longer so.  The reason for using the
absolute value of $S_E$ can be understood by considering the tunneling
wave function for our problem.  To write the corresponding wave
equation, we first substitute (\ref{metric}) into (\ref{action}),
and after integrating by parts find the Lagrangian
\begin{equation}
{\cal L}={1\over{2}}a(1-{\dot a}^2 -H^2a^2 ).
\end{equation}
The momentum conjugate to $a$ is
\begin{equation}
p_a =-a{\dot a},
\label{momentum}
\end{equation}
and the Hamiltonian is
\begin{equation}
{\cal H}=-{1\over{2a}}(p_a^2 +a^2 -H^2a^4).
\label{hamiltonian}
\end{equation}
The evolution equation (\ref{evoleq}) implies that
\begin{equation}
{\cal H}=0.
\end{equation}

Quantization of this model amounts to replacing $p_a \to -i\partial
/\partial a$ and imposing the Wheeler-DeWitt equation
\begin{equation}
{\cal H}\psi =0.
\label{WDW}
\end{equation}
This gives
\begin{equation}
\left[ {d^2\over{da^2}}-U(a)\right]\psi (a)=0,
\label{WDWa}
\end{equation}
where
\begin{equation}
U(a)=a^2(1-H^2a^2),
\label{potential}
\end{equation}
and I have ignored the ambiguity in the ordering of non-commuting
operators $a$ and $p_a$.  (This ambiguity is unimportant in the
semiclassical domain which we will be mainly concerned with in these
lectures).

Eq.(\ref{WDWa}) has the form of a one-dimensional Schrodinger
equation for a `particle' described by a coordinate $a(t)$, having
zero energy, and moving in a potential $U(a)$.  The classically
allowed region is $a\geq H^{-1}$, and the WKB solutions of
(\ref{WDWa}) in this region are
\begin{equation}
\psi_\pm (a)=[p(a)]^{-1/2} \exp \left[ \pm i\int_{H^{-1}}^a p(a')da'
\mp i\pi /4 \right],
\label{psi1}
\end{equation}
where $p(a)=[-U(a)]^{1/2}$.  The under-barrier, $a<H^{-1}$, solutions
are
\begin{equation}
{\tilde \psi}_\pm
(a)=|p(a)|^{-1/2}\exp\left[\pm\int_a^{H^{-1}}|p(a')|da'\right].
\label{psi2}
\end{equation}

For $a\gg H^{-1}$,
\begin{equation}
{\hat p}_a\psi_\pm (a)\approx\pm p(a)\psi_\pm (a),
\end{equation}
and Eq.(\ref{momentum}) tells us that $\psi_-(a)$ and $\psi_+(a)$
describe an expanding and a contracting universe, respectively.  In
the tunneling picture, it is assumed that the universe originated at
small size and then expanded to its present, large size.  This means
that the component of the wave function describing a universe
contracting from infinitely large size should be absent:
\begin{equation}
\psi (a>H^{-1})=\psi_-(a).
\label{tunnel1}
\end{equation}
The under-barrier wave function is found from the WKB connection
formula,
\begin{equation}
\psi (a<H^{-1})={\tilde \psi}_+(a)-{i\over{2}}{\tilde\psi}_-(a).
\label{tunnel2}
\end{equation}
The growing exponential ${\tilde\psi}_-(a)$ and the decreasing
exponential ${\tilde\psi}_+(a)$ have comparable amplitudes at the
nucleation point $a=H^{-1}$, but away from that point the decreasing
exponential dominates (see Fig.~2).  The `nucleation probability' can
be estimated as
\begin{equation}
{\cal P} \sim\exp
\left( -2 \int_0^{H^{-1}}|p(a')|da'\right) = \exp (-|S_E|)=\exp\left(
-{3\over{8G^2\rho_v}}\right).
\label{nuclprob}
\end{equation}
The use of the semiclassical approximation is justified as long as
$|S_E|=3/8G^2\rho_v \gg1$, or $\rho_v \ll \rho_p$, where $\rho_p =
G^{-2}=m_p^4$  is
the Planck density and $m_p$ is the Planck mass.

\begin{figure}
\vspace{5cm}
\caption{\it Tunneling wave function for the de Sitter minisuperspace
model.  The `potential' $U(a)$ is shown by a solid line and the wave
function by a dashed line.}
\end{figure}

Eq.(\ref{nuclprob}) was obtained independently by Linde \cite{Linde84},
Zel'dovich and
Starobinsky \cite{Zelstar}, Rubakov
\cite{Rub84}, and myself \cite{AV84}.  But the story does not end
here.  Not everybody agrees that my first answer (\ref{wrongeq}) was
a mistake.  The same expression for ${\cal P}$ is obtained in the
Euclidean approach developed by Hawking and collaborators.  We shall
return to this on-going debate after discussing the general formalism
of quantum cosmology.
\bigskip

\noindent{\bf 3.\quad Wave Function of the Universe}

\smallskip

\noindent{3.1\quad WHEELER-DE WITT EQUATION}
\medskip
\nobreak

In the general case, the wave function of the universe is defined on
superspace, which is the space of all 3-dimensional geometries and
matter field configurations,
\begin{equation}
\psi [h_{ij}({\bf x}), \varphi ({\bf x})],
\end{equation}
where $h_{ij}$ is the 3-metric, and matter fields are represented by a
single scalar field $\varphi$.  The wave function $\psi$ satisfies the
Wheeler-DeWitt (WDW) equation,
\begin{equation}
{\cal H}\psi (h_{ij},\varphi )=0,
\label{WDWgen}
\end{equation}
which can be thought of as representing the fact that the energy of a
closed universe is equal to zero.  The WDW equation can be
symbolically written in the form
\begin{equation}
(\nabla^2 -U)\psi =0,
\label{Kleingordon}
\end{equation}
which is similar to the Klein-Gordon equation.  Here,
$\nabla^2$ is the superspace Laplacian, and the functional
$U(h_{ij},\varphi)$ can be called `superpotential'.  (We shall not
need explicit forms of $\nabla^2$ and $U$).

We have no idea how to solve the WDW equation in the general case.
Most of what we know about quantum cosmology has been found using
minisuperspace models in which the infinite number of degrees of
freedom in Eq.(\ref{WDWgen}) is reduced to a few independent
variables.  The de Sitter model (\ref{WDWa}) with a single variable
is the simplest example of minisuperspace.  The minisuperspace
approach is justified when the remaining degrees of freedom can be
treated as small perturbations.  The corresponding wave function can
then be calculated perturbatively \cite{Halhaw,VV88}.

Quantum cosmology is based on quantum gravity and shares all of its
problems.  In addition, it has some extra problems which arise when
one tries to quantize a closed universe.  The first problem stems from
the fact that $\psi$ is independent of time.  This can be understood
\cite{Dewitt} in the sense that the wave function of the universe
should describe everything, including the clocks which show time.  In
other words, time should be defined intrinsically in terms of the
geometric or matter variables.  However, no general prescription has
yet been found that would give a function $t(h_{ij},\varphi)$ that
would be, in some sense, monotonic.  A related problem is the
definition of probability.  Given a wave function $\psi$, how can we
calculate probabilities?  One can try to use the conserved current
\cite{Dewitt,Misner}
\begin{equation}
J=i(\psi^*\nabla\psi -\psi\nabla\psi^*),~~~~~~~\nabla\cdot J=0.
\label{current}
\end{equation}
The conservation is a useful property, since we want probability to be
conserved.  But one runs into the same problem as with Klein-Gordon
equation: the probability defined in this way is not
positive-definite.  Although we do not know how to solve these
problems in general, they can both be solved in the semiclassical
domain.  In fact, it is possible that this is all we need.
\bigskip

\noindent{3.2\quad SEMICLASSICAL UNIVERSES}
\medskip
\nobreak

Let us consider the situation when some of the variables $\{ c\}$ describing
the universe behave classically, while the rest of the variables $\{ q \}$ must
be treated quantum-mechanically.  Then the wave function of the universe can be
written as a superposition
\begin{equation}
\psi =\sum_k A_k(c)e^{iS_k(c)}\chi_k (c,q) \equiv\sum_k\psi_k^{(c)}\chi_k,
\label{WKB}
\end{equation}
where the classical variables are described by the WKB wave function
$\psi_k^{(c)}
=A_ke^{iS_k}$.  In the semiclassical regime, $\nabla S$ is large, and
substitution
of (\ref{WKB}) into the WDW equation (\ref{Kleingordon}) yields the
Hamilton-Jacobi equation for $S(c)$,
\begin{equation}
\nabla S\cdot\nabla S +U=0.
\label{hamjac}
\end{equation}
The summation in (\ref{WKB}) is over different solutions of this equation.
Each solution of (\ref{hamjac}) is a classical action describing a congruence
of classical trajectories (which are essentially the gradient curves of $S$).
Hence, a semiclassical wave function $\psi_c =Ae^{iS}$ describes an ensemble of
classical universes evolving along the trajectories of $S(c)$.  A probability
distribution for these trajectories can be obtained using the conserved current
(\ref{current}).  Since the variables $c$ behave classically, these
probabilities
do not change in the course of evolution and can be thought of as probabilities
for various initial conditions.  The time variable $t$ can be defined as any
monotonic parameter along the trajectories, and it can be shown
\cite{Dewitt,AV89} that in this case
the corresponding component of the current $J$ is
non-negative, $J_t \geq 0$.  Moreover, one finds \cite{Laprub,Banks85,Halhaw}
that the `quantum' wave function $\chi$ satisfies the usual Schrodinger
equation,
\begin{equation}
i\partial\chi /\partial t =H_\chi \chi
\end{equation}
with an appropriate Hamiltonian $H_\chi$.  Hence, all the familiar physics is
recovered in the semiclassical regime.

This semiclassical interpretation of the wave function $\psi$ is valid to the
extent that the WKB approximation for $\psi_c$ is justified and the
interference between different terms in (\ref{WKB}) can be neglected.
Otherwise, time and probability cannot be defined, suggesting that the wave
function has no meaningful interpretation.  In a universe where no object
behaves classically (that is, predictably), no clocks can be constructed, no
measurements can be made, and there is nothing to interpret.
\bigskip

\noindent{3.3\quad BOUNDARY CONDITIONS}
\medskip
\nobreak

As (almost) any differential equation, the WDW equation has an infinite number
of solutions.  To get a unique solution, one has to specify some boundary
conditions in superspace.  In ordinary quantum mechanics, the boundary
conditions for the wave function are determined by the physical setup external
to the system under consideration.  In quantum cosmology, there is nothing
external to the universe, and it appears that a boundary condition should
be added to Eq.(\ref{WDWgen}) as an independent physical law.

Several candidates for this law of boundary conditions have been proposed.
Hartle and Hawking \cite{HH83} suggested that $\psi (h,\varphi)$ should be
given
by a path integral over compact, Euclidean 4-geometries $g_{\mu\nu}({\bf
x},\tau)$ bounded by the 3-geometry $h_{ij}({\bf x})$ with the field
configuration $\varphi ({\bf x})$:
\begin{equation}
\psi=\int^{(h,\varphi)}[dg][d\varphi]\exp [-S_E(g,\varphi)].
\label{HHpsi}
\end{equation}
In this path-integral representation, the boundary condition corresponds to
specifying the class of histories integrated over in Eq.(\ref{HHpsi}).  Compact
4-geometries can be thought of as histories interpolating between a point
(`nothing') and a finite 3-geometry $h_{ij}$.

Alternatively, I proposed \cite{AV84,AV85} that $\psi (h,\varphi)$ should be
obtained by integrating over Lorentzian histories interpolating between a
vanishing 3-geometry $\emptyset$ and $(h,\varphi)$ and lying to the past of
$(h,\varphi)$:
\begin{equation}
\psi (h,\varphi)=\int_\emptyset^{(h,\varphi)}[dg][d\varphi]e^{iS}.
\label{lorentzpsi}
\end{equation}
This wave function is closely related to Teitelboim's causal propagator
\cite{Teitel} $K(h_2,\varphi_2 | h_1,\varphi_1)$:
\begin{equation}
\psi(h,\varphi)=K(h,\varphi |\emptyset).
\end{equation}

Linde \cite{Linde84} suggested that, instead of the standard Euclidean rotation
$t \to -i\tau$, the action $S_E$ in (\ref{HHpsi}) should be obtained by
rotating in the opposite sense, $t \to +i\tau$.

Halliwell and Hartle
\cite{Halhar} discussed a path integral over complex metrics which are not
necessarily purely Lorentzian or purely Euclidean.  This encompasses all of the
above proposals and opens new possibilities.  However, the space of complex
metrics is very large, and no obvious choice of integration contour
suggests itself as the preferred one.

In addition to these path-integral no-boundary proposals, one candidate law of
boundary conditions has been formulated directly as a boundary condition in
superspace.  This is the so-called tunneling boundary condition
\cite{AV86,AV88} which requires that $\psi$ should include only outgoing waves
at boundaries of superspace.  It has been argued \cite{AV94} that, in a wide
class of models, this boundary condition is equivalent to the Lorentzian path
integral proposal (\ref{lorentzpsi}).

For the simple de Sitter model of Sec.2, the tunneling wave function
$\psi_T(a)$ is given by Eqs.(\ref{tunnel1}),(\ref{tunnel2}), the Hartle-Hawking
wave function is \cite{Haw84}
\begin{equation}
\psi_H(a>H^{-1})=\psi_+(a)-\psi_-(a),
\end{equation}
\begin{equation}
\psi_H(a<H^{-1})={\tilde \psi}_-(a),
\end{equation}
and the Linde wave function is \cite{Linde84,Lindebook,Slava}
\begin{equation}
\psi_L(a>H^{-1})={1\over{2}}[\psi_+(a)+\psi_-(a)],
\end{equation}
\begin{equation}
\psi_L(a<H^{-1})={\tilde \psi}_+(a).
\end{equation}
Unlike the tunneling wave function, both Hartle-Hawking and Linde wave
functions include expanding and contracting universe components with equal
amplitudes (see Fig.~3).

\begin{figure}
\vspace{8cm}
\caption{\it Hartle-Hawking (a) and Linde (b) wave functions for de Sitter
minisuperspace model.}
\end{figure}
\bigskip

\noindent{\bf 4.\quad Predictions for the Initial State}
\smallskip

\noindent{4.1\quad INITIAL VACUUM ENERGY}
\medskip
\nobreak

To see what kind of cosmological predictions we can get from different boundary
conditions, I would like to consider a somewhat more realistic model.  Instead
of a constant vacuum energy $\rho_v$, I introduce a scalar field $\varphi$ with
a potential $V(\varphi)$.  Since vacuum energy is very small in our part of the
universe, $V(\varphi)$ should have a minimum with $V\approx 0$.  The WDW
equation for this two-dimensional model can be solved assuming that
$V(\varphi)$ is a slowly-varying function and is well below the Planck density,
\begin{equation}
|V'/V|\ll m_p^{-1},~~~~~~~~~~~ V\ll \rho_p \equiv m_p^4.
\label{conditions}
\end{equation}
A slowly-varying $V(\varphi)$ helps to simplify the equation, but is also
necessary for the inflationary scenario.   If the condition $V(\varphi)\ll
\rho_p$ is violated, then the semiclassical approximation is not valid and
higher-order corrections to quantum gravity are important.

After an appropriate rescaling of the scale factor $a$ and the scalar field
$\varphi$, the WDW equation can be written as
\begin{equation}
\left[ {\partial^2 \over{\partial a^2}}-{1\over{a^2}}{\partial^2
\over{\partial\varphi^2}}-U(a,\varphi)\right]\psi(a,\varphi)=0,
\label{WDWaphi}
\end{equation}
where
\begin{equation}
U(a,\varphi)=a^2[1-a^2V(\varphi)].
\label{potentialaphi}
\end{equation}
With the assumptions (\ref{conditions}), one finds \cite{AV88} that
Hartle-Hawking, Linde, and tunneling solutions of this equation are given
essentially by the same expressions as for the simple model (\ref{WDWa}), but
with $\rho_v$ replaced by $V(\varphi)$.  The only difference is that the wave
function is multiplied by a factor $C(\varphi)$, such that
$\psi(a,\varphi)$ becomes
$\varphi$-independent in the limit $a\to 0$ (with $|\varphi |<\infty$).

The initial state of the nucleating universe in this model is characterized by
the value of the scalar field $\varphi$, with the initial value of $a$ given by
$a=V^{-1/2}(\varphi)$.  The probability distribution for $\varphi$ can be found
using the conserved current (\ref{current}),
\begin{equation}
\partial_a J^a +\partial_\varphi J^\varphi =0.
\end{equation}
With a proper normalization, the quantity $\rho(a,\varphi)d\varphi$, where
\begin{equation}
\rho(a,\varphi)=J^a(a,\varphi)=i(\psi^*\partial_a\psi -\psi\partial_a\psi^*),
\end{equation}
can be interpreted as the probability for the scalar field to be between
$\varphi$ and $\varphi +d\varphi$ when the scale factor is equal to $a$.

For the tunneling wave function one finds
\begin{equation}
\rho_T(\varphi)\approx C_T\exp \left( -{3\over{8G^2V(\varphi)}}\right),
\label{tunnelprob}
\end{equation}
where $C_T$ is a normalization constant.  This is the same as
Eq.(\ref{nuclprob}) with $\rho_v$ replaced by $V(\varphi)$.  $\rho$ is
independent of $a$ because $\varphi$ remains approximately constant along the
classical trajectories (with $a$ playing the role of time).  The probability
distribution (\ref{tunnelprob}) is strongly peaked at the value $\varphi
=\varphi_{max}$ where $V(\varphi)$ has a maximum.  Thus, the tunneling wave
function `predicts' that the universe is most likely to nucleate with the
largest possible vacuum energy.  This is just the right initial condition for
inflation.  The high vacuum energy drives the inflationary expansion, while the
field $\varphi$ gradually `rolls down' the potential hill, and ends up at the
minimum with $V(\varphi)\approx 0$, where we are now.

For Linde's wave function, evaluation of the current for the expanding-universe
component of $\psi_L$ gives the same probability distribution
(\ref{tunnelprob}).  The Hartle-Hawking wave function gives a similar
distribution, but with a crucial difference in sign,
\begin{equation}
\rho_H(\varphi)=C_H\exp \left(+{3\over{8G^2V(\varphi)}}\right).
\label{HHprob}
\end{equation}
Note that $\rho_H$ is the same as the nucleation probability (\ref{wrongeq})
found using the instanton method.  This is not surprising: the Hartle-Hawking
proposal involves the same Euclidean rotation as the one used to obtain the
instanton.  The distribution (\ref{HHprob}) is peaked at $V(\varphi)\approx 0$,
and thus the Hartle-Hawking wave function appears to predict an empty universe
with $V\approx 0$.  Such initial condition does not lead to inflation, and is
therefore inconsistent with observations.

Hawking and Page \cite{Hawpage} have pointed out that things may be not so bad
in models of `chaotic' inflation, where $V(\varphi)\to\infty$ at $\varphi \to
\infty$ (a typical example is $V(\varphi) \propto \varphi^{2k}$ with $k$ an
integer).  In such models, $\rho(\varphi\to\infty)\to const$, the distribution
(\ref{HHprob}) is not normalizable, and the ensemble described by this
distribution is dominated by universes with arbitrarily large initial values of
$\varphi$.  The problem with this argument is that, in order to outweigh the
exponentially large values of $\rho_H(\varphi)$ at small $\varphi$, one has to
go to extremely large values of $\varphi$, for which the potential $V(\varphi)$
will far exceed the Planck energy density [except, perhaps, for a very special
shape of $V(\varphi)$].  The semiclassical approximation, on which the
derivation of Eq.(\ref{HHprob}) was based, cannot be trusted in this regime.
For a futher discussion of this issue see Refs. \cite{Grish,Barv}.
\bigskip

\noindent{4.2\quad INITIAL STATE OF THE MODULI}
\medskip
\nobreak

The tunneling {\it vs.} Hartle-Hawking debate takes an interesting turn in
superstring theories, where the relevant part of the low-energy effective
action has the form
\begin{equation}
S=\int d^4 x\sqrt{-g}\left[{R\over{16\pi G}}-K_{AB}(\varphi)g^{\mu\nu}
\partial_\mu\varphi^A\partial_\nu\varphi^B -V(\varphi)\right].
\label{maction}
\end{equation}
Here, $\varphi^A$ are the moduli fields, $K_{AB}(\varphi)$ is the metric of
moduli space, $A,B=1,2,...,n$, and $n$ is the number of moduli.  The
potential $V(\varphi)$ vanishes to all orders of perturbation theory, but is
expected to be generated non-perturbatively, with a characteristic scale well
below the Planck mass.  It has been argued that moduli are natural candidates
for the role of the inflaton in superstring cosmology
\cite{Gaillard,Banks94,Thomas}.

As before, we shall restrict ourselves to a closed Robertson-Walker universe
with homogeneous moduli fields.  Then, after an appropriate rescaling of $t$,
$a$, $\varphi^A$, and $V(\varphi)$, the Lagrangian for our model can be written
as
\begin{equation}
{\cal L}={1\over{2}}a(1-{\dot a}^2)+a^3\left[{1\over{2}}K_{AB}(\varphi) {\dot
\varphi}^A{\dot\varphi}^B -V(\varphi)\right]
\end{equation}
The momenta conjugate to $a$ and $\varphi^A$ are
\begin{equation}
p_a =-a{\dot a}, ~~~~~~~~~~ p_A =a^3K_{AB}{\dot \varphi}^B,
\end{equation}
and the Hamiltonian is
\begin{equation}
{\cal H}={1\over{2a}}[-p_a^2+a^{-2}K^{AB}p_Ap_B-U(a,\varphi)],
\end{equation}
where $U(a,\varphi)$ is given by (\ref{potentialaphi}) and $K^{AB}$ is related
to $K_{AB}$ by the standard relation $K^{AB}K_{BC}=\delta^A_C$.

The WDW equation is obtained by replacing $p_a\to -i\partial /\partial a$,
$p_A\to -i\partial /\partial\varphi_A$,
\begin{equation}
\left[ -{\partial^2\over{\partial a^2}}+{1\over{a^2}}|K|^{-1/2}{\partial\over
{\partial\varphi^A}}\left( |K|^{1/2}K^{AB}{\partial\over{\partial\varphi^B}}
\right)+U(a,\varphi)\right]\psi =0,
\label{mWDW}
\end{equation}
where $K=det(K_{AB})$ and
the ordering of factors $\varphi^A$ and $\partial /\partial\varphi^A$ has
been chosen so that the equation is invariant with respect to
reparametrizations of the moduli space, $\varphi^A\to{\tilde
\varphi}^A(\varphi^B)$ \cite{Ksi}.  The probability distribution for
$\varphi^A$ is then
$d{\cal P}=\rho(a,\varphi)d^n\varphi$, where
\begin{equation}
\rho =J^a=i|K|^{1/2}(\psi^*\partial_a\psi-\psi\partial_a\psi^*).
\end{equation}
For a slowly-varying potential, Eq.(\ref{mWDW}) is solved in the same way as
Eq.(\ref{WDWaphi}) for a single scalar field, and one finds
\begin{equation}
\rho(\varphi)=C|K|^{1/2}\exp\left( \pm{3\over{8G^2 V(\varphi)}}\right),
\end{equation}
where the upper sign corresponds to Hartle-Hawking, and the lower sign to the
tunneling wave function.

Now, the moduli space is non-compact, and one could expect that the remedy
suggested by Hawking and Page \cite{Hawpage} to avoid the empty universe
problem should work in this case as well.  However, Horne and Moore have argued
\cite{Moore} that, despite the existence of non-compact regions, the volume of
the moduli space is finite,
\begin{equation}
\int |K|^{1/2}d^n\varphi <\infty.
\end{equation}
Moreover, it is expected that the moduli potential $V(\varphi)$ asymptotically
vanishes in all non-compact directions.  If either of these expectations is
correct, then the Hartle-Hawking probability distribution is unavoidably peaked
at very low densities, so that
the initial states leading to inflation are highly
unlikely.
\bigskip

\noindent{\bf 5.\quad Predictions for the Constants of Nature}
\smallskip

\noindent{5.1\quad VARIABLE CONSTANTS}
\medskip
\nobreak

In theories that allow variation of the constants of Nature, the cosmological
wave function is a superposition
\begin{equation}
\psi=\sum_\alpha \psi_\alpha(a,\varphi),
\end{equation}
where the
subscript $\alpha$ is a collective symbol for the constants $\{\alpha_j\}$.  In
superstring theories, some of
the constants parametrize different compactifications of
extra dimensions, and different sets of $\{\alpha_j\}$ correspond to
different  moduli
spaces with their own potentials $V(\varphi)$.  The number of different
compactifications is believed to be ${\buildrel > \over \sim} 10^4$.
Hence, the spectrum of $\{\alpha_j\}$ can be rather dense.

The wave function $\psi_\alpha (a,\varphi)$ gives the amplitude for a universe
to nucleate with a set of constants $\{\alpha_j\}$ and to have the values
$a,\varphi$ for the scale factor and moduli fields.  The relative normalization
of different components is fixed for both Hartle-Hawking and tunneling wave
functions if one uses the path integral formulation of the corresponding
boundary conditions, Eqs.(\ref{HHpsi}),(\ref{lorentzpsi}).  The overall
normalization is determined by
\begin{equation}
\sum_\alpha \int\rho_\alpha(\varphi)d\varphi =1.
\end{equation}

We can think of the probability distribution $\rho_\alpha(\varphi)$ as
describing an ensemble of universes which, following Gell-Mann \cite{Gellmann},
I will call `multiverse'.  The probability that a universe arbitrarily picked
in this multiverse will have a particular set of $\{\alpha_j\}$ is
\begin{equation}
w_\alpha =\int\rho_\alpha (\varphi)d\varphi.
\end{equation}
Adopting the tunneling boundary condition for $\psi$, we expect
\begin{equation}
w_\alpha \propto\exp \left(-{3\over{8G^2(\alpha)V_{max}(\alpha)}}\right),
\label{w}
\end{equation}
where $V_{max}(\alpha)=max \{V(\varphi)\}$ for the constants $\{\alpha\}$.

It has been recently suggested \cite{Strom} that most, if not all,
moduli spaces may actually be connected to one another.  The
potential on this interconnected web of moduli spaces may still have a
large number of maxima and minima, with different low-energy physics
at each minimum.  The constants $\{\alpha_j\}$ then parametrize
different minima of $V(\varphi)$, and the probabilities $w_\alpha$ are
obtained by integrating $\rho(\varphi)$ over the basin of attraction
of the corresponding minimum.  We still expect the estimate (\ref{w})
to apply, with $V_{max}(\alpha)$ being the highest maximum of
$V(\varphi)$ in the basin of attraction of the minimum $\alpha$.

The potential $V(\varphi)$ on the moduli space may have some flat
directions.  The associated massless fields may not be in conflict
with observations if they are very weakly coupled.  The values of
these fields, which affect the `constants' of low-energy physics, will
then be determined by the initial conditions at nucleation and by the
following cosmological evolution.  Such fields should be included in
$\{\alpha_j\}$ as continuous variables parametrizing the constants of
Nature \cite{Dilaton}.

Finally, Coleman \cite{Coleman} has argued that all constants
appearing in sub-Planckian physics may become totally undetermined due
to Planck-scale wormholes connecting distant regions of spacetime.
Then the spectrum of the constants is also continuous, but unlike
massless moduli, they cannot vary from one spacetime point to another,
but only from one universe to another (disconnected) universe.
To simplify the discussion, I will assume a discrete spectrum of the
constants.
\bigskip

\noindent{5.2\quad PRINCIPLE OF MEDIOCRITY}
\medskip
\nobreak

It is quite possible that a randomly picked universe will be unsuitable for
life, and therefore the distribution (\ref{w}) is not adequate for predicting
the observed values of the constants.  Moreover, the number of civilizations in
some of the universes may be much greater than in the others, and this
difference should also be taken into account when evaluating
the probabilities \cite{Note2}.
The probability distribution of constants for a {\it civilization} randomly
picked in the multiverse is
\begin{equation}
{\cal P}_\alpha =C^{-1} w_\alpha {\cal N}_\alpha,
\label{p}
\end{equation}
where ${\cal N}_\alpha$ is the average number of civilizations in a universe
with a set of constants $\{\alpha_j\}$ and $C=\sum_\alpha w_\alpha{\cal
N}_\alpha$ is a normalization constant.  ${\cal N}$ is taken to be the total
number of civilizations through the entire history of the universe and is
assumed to be finite.  The case of eternal inflation, where ${\cal N}=\infty$,
will be discussed in Section 6.

If we assume that our civilization is a `typical' inhabitant of the multiverse,
then we `predict' that the constants of Nature in our universe are somewhere
near the maximum of the distribution (\ref{p}).  The assumption of being
typical was called `the principle of mediocrity' in
Ref. \cite{Predictions}.  It
is a version of the `anthropic principle' which has been extensively discussed
in the literature \cite{Anthropic}.

The number ${\cal N}$ can be expressed as
\begin{equation}
{\cal N}_\alpha ={\cal V}_\alpha \nu_{civ}(\alpha),
\label{n}
\end{equation}
where ${\cal V}_\alpha$ is the volume of the universe at the end of inflation
(that is, the 3-volume of the hypersurface that divides the spacetime into
inflating and thermalized parts), and $\nu_{civ}(\alpha)$ is the average number
of civilizations originating per unit thermalized volume.

The definition of probability (\ref{p}) based on the number of
civilizations is somewhat arbitrary.  One could, for example, assign a
weight to each civilization, depending on its lifetime and/or the
number of individuals.  We shall deal with this uncertainty by
concentrating on stable `predictions' from (\ref{p}) which are not
sensitive to the choice of the definition.

The concept of `naturalness' that is commonly used to assess the plausibility
of elementary particle models is based on the assumption that the probability
distribution for the constants is nearly flat,
${\cal P}_\alpha \approx const$.
The principle of mediocrity gives a very different perspective on what is
natural and what is not.  It predicts that the constants $\{\alpha_j\}$ are
likely to be such that the product
\begin{equation}
{\cal P}_\alpha \propto w_\alpha{\cal V}_\alpha \nu_{civ}(\alpha)
\label{p1}
\end{equation}
is maximized.  The factors in this product have a strong (exponential)
dependence on $\{\alpha_j\}$, and the distribution ${\cal P}_\alpha$ can be
strongly peaked in some region of $\alpha$-space.

It should be emphasized that predictions of the principle of
mediocrity are not guaranteed to be correct.  After all, our
civilization may be special in some respects.  The predictions can be
expected to have only statistical accuracy.  That is, with a large
number of predictions, only few of them are likely to be wrong.
\bigskip

\noindent{5.3\quad PREDICTIONS FOR FINITE INFLATION}
\medskip
\nobreak

 From Eq.(\ref{w}), the nucleation probability is maximized when the maximum of
the potential approaches the Planck scale, $V_{max}(\alpha)\sim \rho_p$.  (I
assume that $V(\varphi)$ cannot get much greater than $\rho_p$).

The volume factor ${\cal V}$ is given by ${\cal V}={\cal V}_0 Z^3$,
where ${\cal V}_0 \sim (GV_{max})^{-3/2}$ is the initial volume at
nucleation and $Z$ is the expansion factor during inflation.
The maximum of $Z$ is achieved by maximizing the highest value of the
potential $V_{max}$, where inflation starts, and minimizing the slope
of $V(\varphi)$: the field $\varphi$ takes longer to roll down for a
flatter potential.

The cosmological literature abounds with remarks on the `unnaturally'
flat potentials required by inflationary scenarios.  With the
principle of mediocrity the situation is reversed: flat is natural.
Instead of asking why $V(\varphi)$ is so flat, one should now ask why
it is not flatter.

The `human factor' $\nu_{civ}(\alpha)$ may impose stringent
constraints on the constants $\{\alpha_j\}$.  We do not know what
other forms of intelligent life are possible, but the principle of
mediocrity favors the hypothesis that our form is the most common in
the multiverse.  The conditions required for life of our type to exist
[the low-energy physics based on the symmetry group $SU(3)\times
SU(2)\times U(1)$, the existence of stars and planets, supernova
explosions] may then fix, by order of magnitude, the values of the
fine structure constant, and of electron, nucleon, and W-boson masses,
as discussed in Ref. \cite{Anthropic}.  Anthropic considerations also
impose a bound on the allowed flatness of the potential $V(\varphi)$.
If it is too flat, then the thermalization temperature after inflation
is too low for baryogenesis.  The lowest temperature at which
baryogenesis can still occur is set by the electroweak scale,
$T_{min}\sim m_W$.  Hence, if other constraints do not interfere, we
expect the universe to thermalize at $T\sim m_W$.

Superflat potentials required by the principle of mediocrity typically
give rise to density fluctuations which are many orders of magnitude
below the strength needed for structure formation.  This means that
the observed structures must have been seeded by some other mechanism.
An alternative mechanism is based on
topological defects: strings, global monopoles, and textures, which
could be formed at a symmetry breaking phase transition \cite{Book}.
The required symmetry breaking scale for the defects is $\eta\sim
10^{16}~GeV$.  With `natural' (in the traditional sense) values of the
couplings, the transition temperature is $T_c\sim\eta$, which is much
higher than the  thermalization temperature ($T_{th}\sim m_W$), and no
defects are formed after inflation.  It is possible for the phase
transition to occur during inflation, but the resulting defects are
inflated away, unless the transition is sufficiently close to the end
of inflation.  To arrange this requires some fine-tuning of the
constants.  However, the alternative is to have thermalization at a
much higher temperature and to cut down on the amount of inflation.
Since the dependence of the volume factor ${\cal V}$ on the duration
of inflation is exponential, we expect that the gain in the volume
will more than compensate for the decrease in `$\alpha$-space' due to
the fine-tuning.  We note also that in some supersymmetric models the
critical temperature of superheavy string formation can `naturally' be
as low as $m_W$ \cite{Shafi}.

Another possibility is to use more complicated models of inflation, such as
`hybrid' inflation \cite{Hybinf}, which involve several scalar fields and can
give reasonably large density fluctuations even when the potentials are very
flat in some directions in the field space \cite{Thanklin}.  The amount of
fine-tuning required in these models appears to be comparable to that in the
case of topological defects.

The symmetry breaking scale $\eta\sim 10^{16}~GeV$ for the defects is
suggested by observations, but we have not explained why this
particular scale has been selected.  The value of $\eta$ determines
the amplitude of density fluctuations, which in turn determines the
time when galaxies form, the galactic density, and the rate of star
formation in the galaxies.  Since these parameters certainly affect
the chances for civilizations to develop, it is quite possible that
$\eta$ is significantly constrained by the anthropic factor
$\nu_{civ}(\alpha)$.  It would therefore be interesting to study how
structure formation would proceed in a universe with a very different
amplitude of density fluctuations (and a very different baryon
density).  Some  steps in this
direction have been made in Ref. \cite{Rees}.

If $\nu_{civ}$ is indeed sharply peaked at some value of $\eta$ and
thus fixes the amplitude of density fluctuations and the epoch of
active galaxy formation, then an upper bound on the cosmological
constant can be obtained by requiring that it does not disrupt galaxy
formation until the end of that epoch.  An anthropic bound on the cosmological
constant has been first discussed by Weinberg \cite{Weinberg}.  He argued that,
since there is evidence for the existence of some quasars and protogalaxies as
early as $z \sim 4$, the anthropic principle cannot rule out vacuum energy
domination at $z {\buildrel < \over \sim} 4$.  The matter density at $z = 4$ is
greater than the present matter density $\rho_{mo}$ by a factor $(1+z)^3=125$,
and he concluded that the anthropic bound on $\rho_v$ cannot be stronger than
$\rho_v/\rho_{m0} {\buildrel < \over \sim} 100$.  This falls short of the
observational upper bound \cite{Carroll}, $|\rho_v/\rho_{m0}|_{obs} {\buildrel
< \over \sim} 10$, by a factor $\sim 10$ \cite{Eta}.

On the other hand, the principle of mediocrity suggests that we look not for
the value of $\rho_v$ that makes galaxy formation barely possible, but for the
value that maximizes the amount of matter in galaxies \cite{F}.  This amount
grew substantially after $z=4$, and it is quite possible that it increased,
say, by a factor $\sim 2$ as late as $z \sim 1$.  Requiring that $\rho_v$ does
not dominate before $z \sim 1$, we obtain $\rho_v/\rho_{m0} {\buildrel < \over
\sim} 10$.  The actual value of
$\rho_v$ is likely to be comparable to this upper bound.  Negative
values of $\rho_v$ are bounded by requiring that our part of the
universe does not recollapse while stars are still shining and new
civilizations are being formed.  This gives a bound comparable to that
for positive $\Lambda$ (by absolute value).  A more detailed discussion of the
bounds on the cosmological constant will be given elsewhere.

Let us now summarize the `predictions' of the principle of mediocrity
for the case of finite inflation \cite{Predictions}.
The preferred models have very flat
inflaton potentials, thermalization and baryogenesis at the
electroweak scale, a non-negligible cosmological constant, and density
fluctuations seeded either by topological defects, or by quantum fluctuations
in models like hybrid inflation (as long as these features are
consistent with the spectrum of the constants $\{\alpha_j\}$).
\bigskip

\noindent{\bf 6.\quad Predictions for Eternal Inflation}
\medskip
\nobreak

I have assumed so far that inflation has a finite duration, so that
the thermalized volume ${\cal V}$ and the number of civilizations
${\cal N}$ are both finite.  This, however, is not generally the case.
The evolution of the inflaton field $\varphi$ is influenced by quantum
fluctuations, and as a result thermalization does not occur
simultaneously in different parts of the universe.  In many models it
can be shown that at any time there are parts of the universe that are
still inflating \cite{AV83,Linde86,Topinf}.  The conclusions of
Section 5.3 are directly applicable only if inflation is finite for all
the allowed values of the constants $\{\alpha_j\}$.  For eternally
inflating universes the situation is substantially more complicated.
This subject is now under active investigation, and I will have time
only for a quick review.
\bigskip

\noindent{6.1\quad DISCONNECTED UNIVERSES}
\medskip
\nobreak

Let us first suppose that different sets of constants $\{\alpha_j\}$
correspond to different, disconnected universes.  In order to
calculate the probabilities (\ref{p}), we should then be able to
compare the thermalization volumes ${\cal V}_\alpha$.

In an eternally inflating universe, the thermalization volume ${\cal
V}$ is infinite and has to be regulated.  If one simply cuts it off by
including only parts of the volume that thermalized prior to some
moment of time $t_c$, with the same value of $t_c$ for all universes,
then one finds that the results are extremely sensitive to the choice
of the time coordinate $t$.  For example, cutoffs at a fixed proper
time and at a fixed scale factor $a$ give drastically different
results \cite{LLM}.  An alternative procedure \cite{AV95} is to
introduce the cutoff at the time when all but a small fraction
$\epsilon$ of the initial (co-moving) volume of the universe has
thermalized.  The value of $\epsilon$ is taken to be the same for all
universes, but the corresponding cutoff times $t_c$ are generally
different.  The limit $\epsilon \to 0$ is taken after calculating the
probability distribution ${\cal P}_\alpha$.  It was shown in
\cite{AV95} that the resulting distribution is not sensitive to
the choice of $t$.

The regularized volume ${\cal V}$ can be calculated in terms of the
distribution function ${\cal P}(\varphi,a)$, which is defined so that
${\cal P}(\varphi,a)d\varphi$ gives the fraction of the co-moving
volume where the scalar field(s) takes values between $\varphi$ and
$\varphi +d\varphi$, with the scale factor $a$ playing the role of a time
coordinate.  The function ${\cal P}(\varphi,a)$ satisfies a
`diffusion' equation \cite{AV83,Starob}, which I will not reproduce here.
The important thing for us to know is that the asymptotic form of
${\cal P}$ at large $a$ is
\begin{equation}
{\cal P}(\varphi,a\to\infty)\approx f(\varphi)a^{-\gamma}.
\label{asymptotic}
\end{equation}
The positive constant $\gamma$ can be found by solving an eigenvalue
problem \cite{Starob}, and $d=3-\gamma$ has the meaning of the fractal
dimension of the inflating region \cite{Aryal}.  (It can be shown that
$\gamma\leq 3$).  I will omit the calculations performed in
Ref. \cite{AV95} and even the rather lengthy expression for ${\cal V}$
obtained as a result of those calculations.  The essence of the result
can be expressed as
\begin{equation}
{\cal V}\propto\epsilon^{-(3-\gamma)/\gamma}Z^3.
\label{vol}
\end{equation}
Here, $Z$ is the expansion factor during the slow-roll phase of
inflation, when quantum fluctuations are small.

In the limit $\epsilon\to 0$, non-vanishing probabilities are
obtained only for $\{\alpha_j\}$ corresponding to the smallest value
of $\gamma$,
\begin{equation}
\gamma(\alpha)=min.
\label{gamma}
\end{equation}
The eigenvalue $\gamma$ decreases as the potential $V(\varphi)$
becomes flatter \cite{Aryal,LLM}, and thus the condition (\ref{gamma})
tends to select maximally flat potentials.

It is possible that the condition (\ref{gamma}) selects a unique set
of $\{\alpha_j\}$.  Then all constants of Nature can, at least in
principle, be predicted with 100\% certainty.  On the other hand, it
is conceivable that the minimum of $\gamma$ is strongly degenerate, so
that Eq.(\ref{gamma}) selects a large subset of all $\{\alpha_j\}$.
Then all values of $\alpha$ not in this subset have a vanishing
probability, and the probability distribution within the subset is
proportional to $w_\alpha Z^3_\alpha\nu_{civ}(\alpha)$ [see
Eq.(\ref{p1})].  The probability maximum is then determined by the same
considerations as in the case of finite inflation.

It should be emphasized that the conditions of minimizing $\gamma(\alpha)$ and
maximizing $w_\alpha Z_\alpha^3\nu_{civ}(\alpha)$ are not on an equal footing,
with the first of these conditions always taking precedence.  Even a tiny
decrease in $\gamma$ leads to an infinite increase of the thermalization volume
in the limit $\epsilon \to 0$.  Suppose, for example, that we have two sets of
constants, $\{\alpha_j^{(1)}\}$ and $\{\alpha_j^{(2)}\}$, such that
$\gamma(\alpha^{(1)})<\gamma(\alpha^{(2)})$, but the thermalization temperature
for  the constants $\alpha^{(1)}$ is too low for baryogenesis, while for
$\alpha^{(2)}$ it is sufficiently high.  We would still have to conclude that
the constants $\alpha^{(1)}$ are infinitely more probable than $\alpha^{(2)}$.
In a universe described by the constants $\alpha^{(1)}$, life can appear only
as a result of a huge fluctuation of the baryon density.  The probability of
such a fluctuation per unit volume is incredibly small, but its
smallness is more than
compensated for when the volume is increased by an infinite factor.

\bigskip

\noindent{6.2\quad MULTIPLE VACUA IN A SINGLE UNIVERSE}
\medskip
\nobreak

Let us now consider eternal inflation in a single universe where the potential
$V(\varphi)$ has a large number of minima, parametrized by the constants
$\{\alpha_j\}$.  Thermalization will then ocur in different minima in different
parts of the universe.  The asymptotic form of the distribution function
${\cal P}(\varphi,a)$ in this case is still
given by Eq.(\ref{asymptotic}), but now $\gamma$ has the same value everywhere
and is independent of $\{\alpha_j\}$ \cite{LLM}.
{}From Eq.(\ref{vol}), the regularized thermalization
volumes are ${\cal V}_\alpha \propto Z_\alpha^3$, and the
corresponding probabilities are
\begin{equation}
{\cal P}_\alpha \propto Z_\alpha^3\nu_{civ}(\alpha).
\label{pcms}
\end{equation}

The probability distribution (\ref{pcms}) has the same dependence on
the slow-roll expansion factor $Z$ and on the anthropic factor
$\nu_{civ}$ as we found in the case of finite inflation.
The predictions for $\{\alpha_j\}$ are, therefore, also the same
(see Section 5.3).
\bigskip

\noindent{6.3\quad ETERNAL INFLATION AND QUANTUM COSMOLOGY}
\medskip
\nobreak

The ideas of eternal inflation and quantum cosmology have always had a somewhat
uneasy coexistence.  The picture of an eternally inflating universe, with new
islands of thermalization constantly being formed, makes one wonder about the
possibility of extending this picture to the infinite past.  The universe would
then be in a steady state of eternal inflation without a beginning, the problem
of the initial singularity would be avoided, and there would be no need for
quantum cosmology.  However, it has been shown \cite{Borde} that, under rather
general assumptions, inflation cannot be eternal in both future and past
directions.  Hence, an eternally inflating universe must still have a
beginning, and we probably need quantum cosmology to describe it.

On the other hand, the eternal nature of inflation can make the initial state
at the nucleation of the universe completely irrelevant.  The universe
eventually reaches the steady-state regime described by the asymptotic form
(\ref{asymptotic}) and stays in this regime thereafter.  If transitions between
vacua with different $\{\alpha_j\}$ are in principle possible, then the
universe completely forgets its initial conditions.  In this case, all one
needs from the cosmological wave function is that the probability for eternal
inflation to start should be non-zero.  Hartle-Hawking and tunneling wave
functions are then in equally good agreement with observations, and the
probability distribution for $\{\alpha_j\}$ is given by
Eq.(\ref{pcms}).  (This equation was derived without relying on
quantum cosmology).

Thus, we see that eternal inflation is quite capable of determining
the values of $\{\alpha_j\}$ on its own, without any help from quantum
cosmology.  The inverse is also true: in models where inflation is
finite for all allowed values of $\{\alpha_j\}$, quantum cosmology can
determine the probability distribution for the initial states and for
the constants of Nature, without any need for eternal inflation
\cite{Combined}.  At this time it is hard to tell which of the two
approaches is more promising, and both are probably worth pursuing.
\bigskip

\noindent{Acknowlegements}
\medskip

It is a pleasure to thank the organizers of the course for their warm
hospitality.  I am also grateful to Arvind Borde, Allen Everett, Cumrun Vafa,
Serge Winitzki, and particularly Andrei Linde and Don Page for
discussions and comments.
This work was supported in part by the National Science Foundation.


\begin{thebibliography}{11}
\bibitem{Mistake}
We note that calling something the greatest mistake of one's life may be a
mistake.  For example, the introduction of the cosmological constant, which
Einstein called the greatest mistake of his life, now appers to be not such a
bad idea.
\bibitem{Ksi}
The most general factor ordering consistent with reparametrization
invariance allows an extra term $\xi a^{-2}{\cal R}(\varphi)$ in
Eq.(\ref{mWDW}).  Here, ${\cal R}$ is the scalar curvature of the
moduli space and $\xi$ is a numerical coefficient.  Addition of such a
term modifies the potential $U(a,\varphi)$ but does not change the
conclusions of Sec.4.2.
\bibitem{Dilaton}
The probability distribution for a Brans-Dicke field (which is similar to the
dilaton of superstring theories) was discussed, using a different approach, by
Garcia-Bellido and Linde \cite{GBL,GBL1}.
\bibitem{Eta}
This argument assumes that the probability distribution for $\rho_v$ in the
range of interest  is nearly flat.  It is possible, however, that the
`fundamental' variable that has a flat distribution at sub-Planckian scales is
the characteristic energy scale $\eta = \rho_v^{1/4}$.  Then the discrepancy
between the anthropic and observational bounds on $\eta$ is only by a factor
$\sim 2$.
\bibitem{F}
More exactly, we look for the values of $\rho_v$ that achieve a balance between
fine-tuning and maximizing the amount of matter in galaxies.  To make this
quantitative, let $w(\rho_v)d\rho_v$ be the probability distribution for
$\rho_v$ for the nucleating universes, and let $f(\rho_v)$ be the fraction of
baryonic matter that ends up in galaxies for a given value of $\rho_v$.  (Here
I assume that $\rho_v$ has a continuous spectrum).  Then we want to maximize
the product $f(\rho_v)w(\rho_v)\rho_v$.
\end{thebibliography}
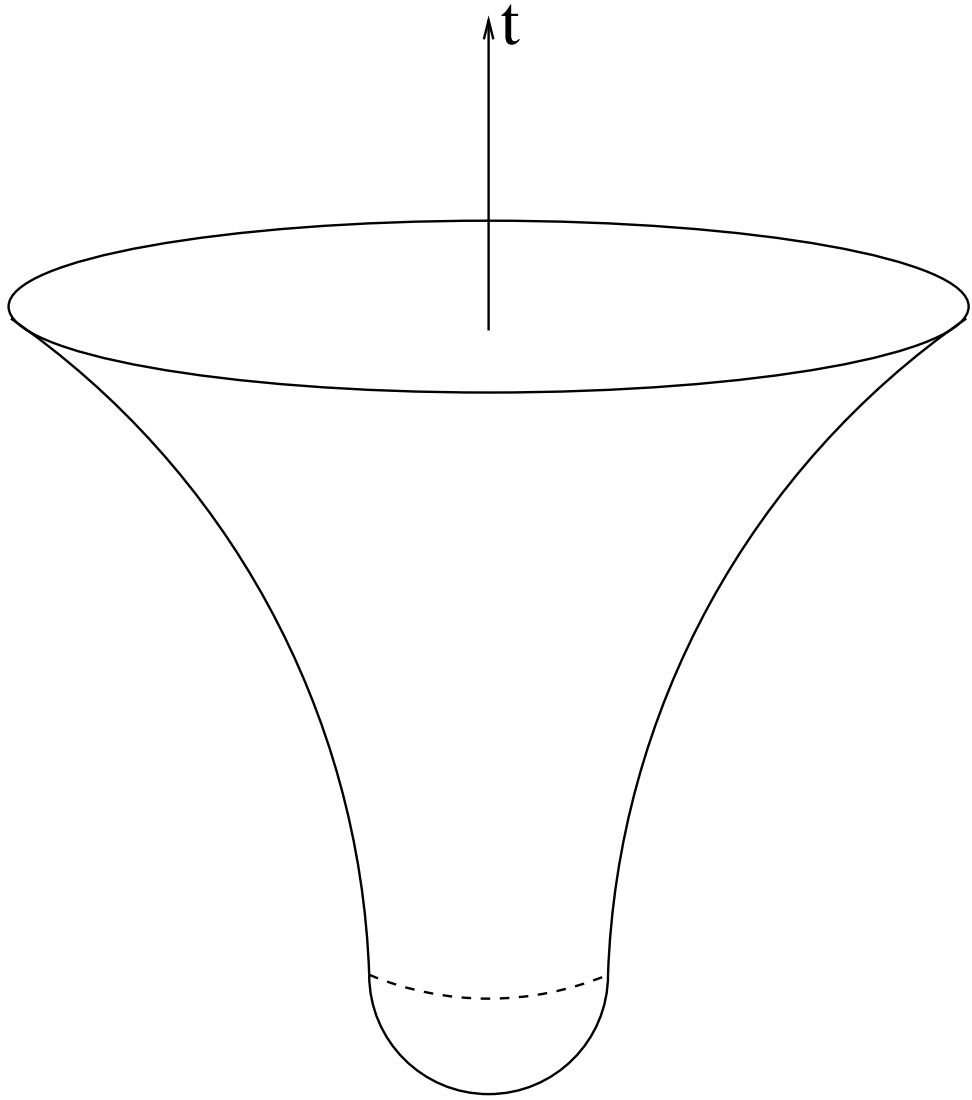
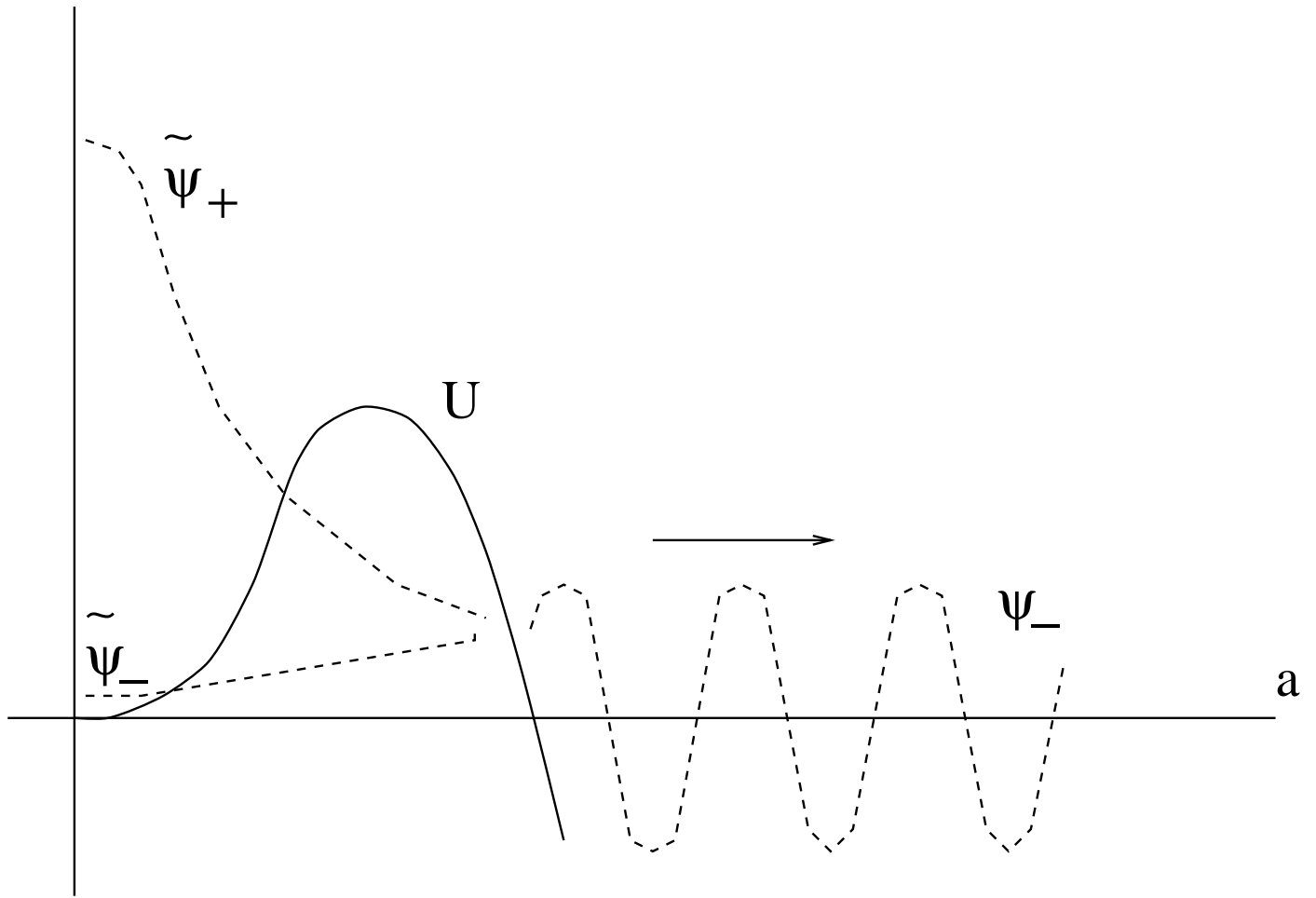
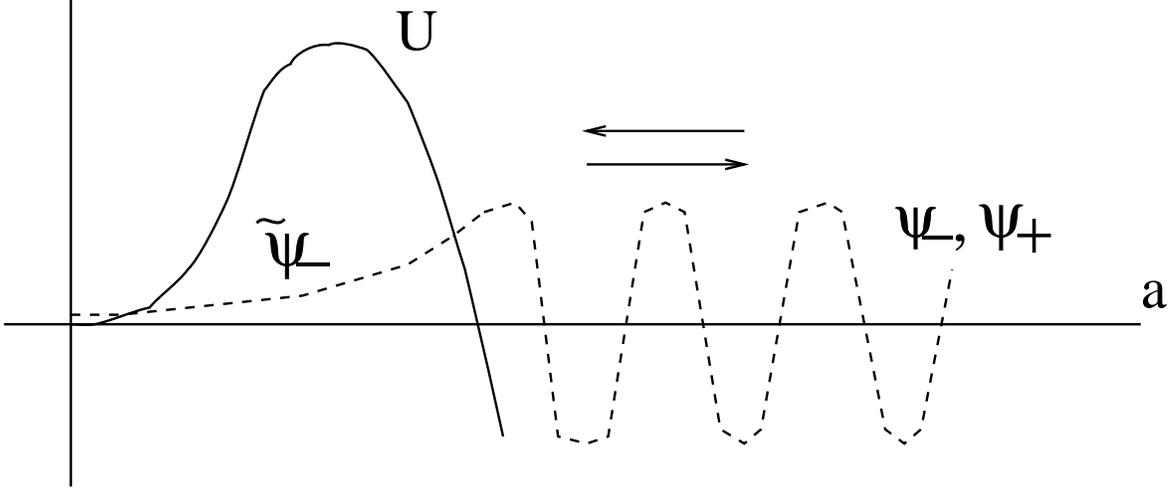
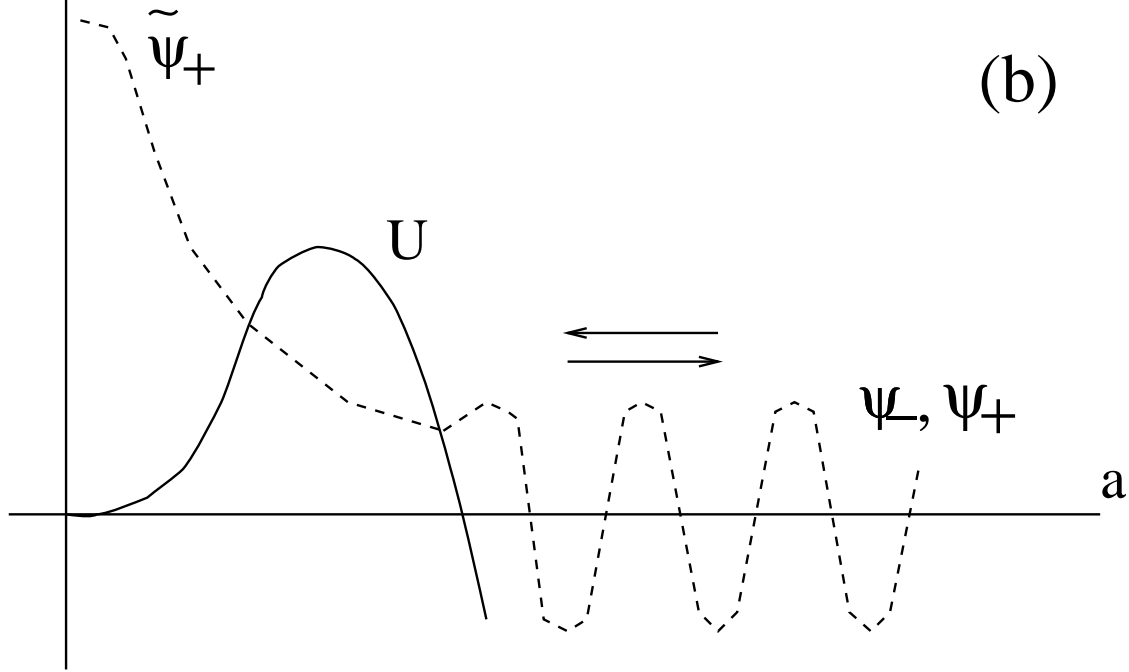
\end{document}